\documentclass{cimento}

\newcommand\be{\begin{equation}}
\newcommand\ba{\begin{eqnarray}}
\newcommand\ee{\end{equation}}
\newcommand\ea{\end{eqnarray}}

\title{String Cosmology and the Breakdown of Local Effective Field Theory}
\author{Robert Brandenberger\\
Physics Department, McGill University, Montreal, Canada
\thanks{Email: rhb@physics.mcgill.ca}}

\begin{document}

\maketitle

\begin{abstract}
Most current cosmological models of the very early universe are based on local point particle effective field theories coupled to gravity. I will discuss some conceptual limitations of this approach and argue that an improved description of the early universe needs to go beyond this framework. I will outline a couple of ideas based on superstring theory.  
\end{abstract}

\section{Introduction}

Standard Big Bang (SBB) cosmology is unable to explain most of the data which modern telescopes are providing us with. Firstly, since the part of the last scattering surface which intersects our past light cone and over which we observe isotropy of the Cosmic Microwave Backgound (CMB) has a radius which is much larger than the horizon at the time of last scattering, causality implies that it is impossible to explain the observed isotropy. Similarly, since comoving scales on which we currently map out the nonrandom structure in the distribution of galaxies and the CMB anisotropies have a wavelength larger than the horizon at the time of equal matter and radiation, it is not possible to in a causal way explain the origin of the nonrandom distribution of seed fluctuations which evolve by gravity into the currently observed structure. In order to solve these problems of the SBB, we need to modify the evolution of space-time in the very early universe.

There are three criteria which a successful early universe cosmology must satisfy. Firstly, in order to allow for a causal explanation of the isotropy of the CMB, the horizon (defined as the forward light cone of an initial time Cauchy surface) has to be much larger than what it appears to be in the SBB scenario. Secondly, in order to allow for a causal generation mechanism of the fluctuations which evolve into the observed structure of the universe today, the comoving scale of these fluctuations must be smaller than the Hubble radius $H^{-1}(t)$ (where $H$ is the expansion rate of space) at some early time. Finally, there needs to be a causal mechanism which produces a nearly scale-invariant spectrum of nearly Gaussian and nearly adiabatic fluctuations. If these criteria are satisfied, then, as discussed in \cite{SZ} and \cite{PY} already ten years before the development of inflationary cosmology, acoustic oscillations in the angular power spectrum of the CMB and baryon acoustic oscillations (oscillations in the power spectrum of matter perturbations) will be produced.

{\it Cosmological inflation} \cite{Guth} is one scenario which satisfies the three abovementioned criteria. During the period of exponential expansion of space which inflationary cosmology postulates, the horizon expands exponentially relative to the Hubble radius. The same exponential expansion of space ensures that, provided that the period of inflation is sufficiently long, scales which are probed now in cosmological observations originate at the beginning of the period of inflation on sub-Hubble scales. Finally, the translation-invariance of the local physics during the phase of inflation ensures \cite{Press} that any structure formation mechanism, be it quantum vacuum perturbations \cite{ChibMukh} in canonical slow-roll inflation models or thermal fluctuations in the warm inflation scenario \cite{Berera} will lead to an almost scale-invariant spectrum of curvature perturbations \cite{ChibMukh} and gravitational waves \cite{Starob}. 

However, inflation is not the only viable scenario. {\it Bouncing cosmologies} can trivially satisfy the first two criteria: if time is past eternal, the horizon is infinite, allowing a causal explanation of the isotropy of the CMB. In addition, all scales are sub-Hubble at sufficiently early times during the contracting phase, thus making it possible to have a causal generation mechanism for fluctuations. As pointed out in \cite{Fabio}, if fluctuations originate as quantum vacuum perturbations, then, on scales which exit the Hubble horizon during a matter-dominated phase of contraction, a scale-invariant spectrum of both curvature perturbations and gravitational waves will be produced. There are two major challenges for such a ``matter bounce'' scenario, the first being the instability to the growth of anisotropies \cite{Yifu}, the other the difficulty in obtaining a sufficiently small tensor-to-scalar ratio (ratio of the power spectra of gravitational waves and curvature fluctuations) while not generating too large non-Gaussianities \cite{Jerome}. A scenario with a very slow contracting phase \cite{Ekp} is more promising since such a phase dilutes anisotropies, creates spatial flatness and is an attactor in initial condition space \cite{Stein}. In a recent realization of the scenario in which an S-brane mediates the transition between the contracting and expanding phases, a scale-invariant spectrum of both curvature fluctuations and gravitational waves can be obtained \cite{Ziwei}.

A scenario which will be relevant to the later discussion of superstring cosmology is the {\it emergent scenario} where it is assumed that the expanding phase of Standard Big Bang cosmology emerges from an early and possibly non-geometric phase at the end of which we have thermal fluctuations with holographic scaling. In an early realization \cite{BV}, the emergent phase was modelled as a quasi-static phase of a gas of strings close to the Hagedorn temperature \cite{Hagedorn}, the maximal temperature of a gas of closed strings. As in bouncing scenarios, the horizon is infinite, and scales which are mapped out today originate sub-Hubble during the emergent phase. Furthermore, the holographic scaling of the primordial curvature fluctuations and gravitational waves is a direct consequence of the fact that the basic degrees of freedom are strings and not point particles \cite{NBV}.

The standard realizations of the inflationary scenario are based on an effective field theory description of matter, as are a number of the proposed bouncing scenarios. In the following I will discuss some conceptual challenges which an effective field theory analysis of early universe cosmology faces, with particular emphasis on how the cosmological fluctuations are described.


\section{Effective Field Theory Analysis of Cosmological Fluctuations}

Curvature fluctuations and gravitational waves must be very small in amplitude in the early universe and can hence be well described in terms of the theory of linearized cosmological perturbations (see e.g. \cite{MFB, RHBfluctrev} for reviews). In longitudinal gauge, and in the case of matter without anisotropic stress, the space-time metric takes the form
\be
ds^2 \, = \, a(\eta)^2 \bigl( (1 + 2 \Phi) d\eta^2 - [(1 - 2\Phi)\delta_{ij} + h_{ij}]dx^idx^j \bigr) \, ,
\ee
where $a(\eta)$ is the scale factor, $\eta$ is conformal time which is related to the physical time $t$ via $dt = a(\eta)d\eta$, $\Phi(x,\eta)$ describe the curvature perturbations, and $h_{ij}(x, \eta)$ is the gravitational wave tensor. Latin indices are used for the comoving spatial coordinates $x^i$.

The action for linear cosmological perturbations is the sum of an action for the curvature perturbations and an action for gravitational waves. The action for cosmological perturbations takes the canonical form in terms of a rescaled field $v$ which, in the case of matter being a canonically normalized scalar field $\varphi$ with a time dependent background $\varphi_0(\eta)$ takes the form \cite{Mukh, Sasaki}
\be
v \, = \, a \bigl[ \delta \varphi + \frac{\varphi_0^{\prime}}{\cal{H}} \Phi \bigr] \, ,
\ee
where a prime denotes the derivative with respect to $\eta$, and ${\cal{H}}$ is the Hubble expansion rate in comoving time, i.e. ${\cal{H}} = a^{\prime} / a$.

At the linearized level, each comoving spatial Fourier mode of the fluctuations evolves independently, and the equation of motion for the mode with comoving wavenumber $k$ takes the form
\be \label{fluct}
v_k^{\prime \prime} + \bigl[ k^2 - \frac{z^{\prime \prime}}{z} \bigr] v_k \, = \, 0 \, ,
\ee
where $z = a \varphi_0^{\prime} / {\cal{H}}$ is a function of the background which is proportional to $a$ if the equation of state of the background is independent of time. The equation of motion for the rescaled gravitational wave ampliutde is the same, except that the function $z(\eta)$ is now simply $a(\eta)$.

For relativistic matter, the magnitude of $z^{\prime \prime} / z$ is given by ${\cal{H}}^2$. Hence, it follows from (\ref{fluct}) that on sub-Hubble scales $k > {\cal{H}}$ the canonical fluctuation variables simply oscillate, while on super-Hubble they freeze out, are squeezed, and can classicalize as a consequence of decoherence and nonlinearities \cite{Kiefer}. These features will play an important role in the following discussion.

\section{Challenges for an Effective Field Theory Description of Early Universe Cosmology}

Effective field theory (see e.g. \cite{EFTrevs} for reviews) is based on the idea that the low energy physics can be well desribed by a field theory resulting from integrating out high energy modes. In standard particle physics this method is applied widely and very successful. However, field theory in an expanding background provides a challenge to the applicability of effective field theory. The reason is that each degree of freedom of the effective field theory is a harmonic oscillator mode with fixed {\it comoving} wavenumber. To tame the divergences which arise in any point particle field theory, a cutoff must be introduced, and this cutoff corresponds to a fixed {\it physical} length scale. As already pointed out in \cite{Weiss}, in order to maintain this ultraviolet cutoff, field modes need to be continuously created. This is a violation of perturbative unitarity. A conservative demand for a consistent effective field theory description of the early universe is that no mode which initially has an energy in excess of the cutoff scale become super-Hubble. In this case the non-unitarity does not effect fluctuation modes which get squeezed and can classicalize \cite{RHBrev}. The mathematical form of this criterion \cite{TCC1} is that the comoving length scale associated with the cutoff energy (here taken to be the Planck length $l_{pl}$) at some initial time $t_i$ should never grow to become larger than the Hubble horizon radius at some later time $t_R$, i.e.
\be \label{TCC}
\frac{a(t_R)}{a(t_i} l_{pl} \, \leq \, H^{-1}(t_R) \, .
\ee
In cosmologies in which this criterion is not met, a {\it trans-Planckian problem} for fluctuations arises \cite{MB}: new physics beyond what is contained in the effective field theory must enter in order to determine the initial conditions for the fluctuations. In particular, in the case of inflationary cosmology, if the period of inflation lasts only slightly longer than it has to in order that scales which are currently mapped out in experiments originate sub-Hubble at the beginning of the period of inflation, then these modes originate in the trans-Planckian sea of ignorance. In the case of cosmological inflation, (\ref{TCC}) leads to an upper bound on the number $N$ of e-foldings of exponential expansion
\be \label{TCCN}
N \, < \, {\rm{ln}} \frac{m_{pl}}{H} \, ,
\ee
where $m_{pl}$ is the Planck mass.

A second way at arriving at the {\it Trans-Planckian Censorship Conjecture} (\ref{TCC})(TCC) is as an analogy to Penrose's Cosmic Censorship Hypothesis \cite{Penrose} which states that timelike singularities need to be hidden by horizons in a consistent theory of quantum gravity. For example, while General Relativity as a low energy effective field theory admits black hole soluitons with charge $Q$ greater than the mass $M$, solutions which have naked singularities, a full quantum theory of gravity should prohibit such solutions. The observer outside of the black hole horizon should be shielded from the non-unitarity associated with the black hole singularity. As already discussed in \cite{RHBrev}, if we adopt the translation scheme where position space for black holes is mapped to momentum space for cosmology, the black hole singularity becomes the sea of trans-Planckian modes, and the black hole horizon is replaced by the Hubble horizon, we obtain the conjecture that in any consistent quantum theory of gravity, observers who have access to super-Hubble modes should never be able to access modes which at some point in time were in the trans-Planckian sea. This yields precisely the condition (\ref{TCC}). Effective field theories which violate this conjecture will hence be inconsistent with quantum gravity. 

The reason for focusing on the Hubble horizon as the analog of the black hole horizon in the above argument is that, as discussed in the previous section, fluctuations only squeeze and become classical on super-Hubble scales. The reason for using the Hubble scale in the TCC is that we do not want fluctuations which grow and become classical to be effected by trans-Planckain effects.

In cosmological models without an extended phase of accelerated expansion, the TCC condition (\ref{TCC}) does not lead to any constraints since in this case the Hubble radius expands faster that the physical wavelength of any fixed comoving scale. However, in the case of inflationary cosmology the TCC leads to stringent constraints \cite{TCC2}: the TCC criterion imposes an upper bound on the duration of inflation, while inflation being successful at providing a causal mechanism of generating fluctuations on the largest observed cosmological scales leads to a lower bound. The upper bound states that the initial Hubble radius must remain sub-Hubble at the end of inflation, while the lower bound states that the comoving scale corresponding to the current Hubble radius must originate with a length smaller that the Hubble radius at the beginning of inflation. A simple calculation shows that, for almost exponential inflation and for rapid reheating (energy transfer within one Hubble time at the end of inflation), the two conditions can only be met for very low scale inflation:
\be \label{bound}
\eta \, < \, 3 \times 10^{9} {\rm{GeV}} \, 
\ee
(compared to the scale of about $10^{15} {\rm{GeV}}$ which is required in canonical slow-roll inflation models without fine tunings of the matter sector). For such low scale inflation the predicted amplitude of primordial gravitational waves is negligible compared to waves induced by secondary effects. Note that the bound (\ref{bound}) can be strengthened by taking into account the pre-inflationary radiation phase \cite{Ed}, but it can be relaxed for power law inflation, and for slow reheating models (see e.g. \cite{Kamali}).

A third way of justifying the TCC condition (\ref{TCC}) is by studying the entanglement entropy between sub- and super-Hubble modes induced by the leading nonlinearities (the pure entanglement entropy due to squeezing of the super-Hubble modes was already considered in \cite{Tomislav}). In the case of inflationary cosmology, demanding that the entanglement entropy not exceed the thermal entropy after inflation \cite{Suddho1} or the Gibbons-Hawking entropy of a de Sitter phase \cite{Suddho2} yields an upper bound on the number of e-foldings which, up to a numerical factor, agrees with (\ref{TCCN}). This analysis is closely related to work which shows that de Sitter space is a ``fast scrambler'' (see e.g. \cite{Shiu}).

The TCC is just one manifestation of the breakdown of point particle-based local effective field theory in cosmology. Another manifestation is the cosmological constant problem, namely the divergence of the effective field theory vacuum energy as the ultraviolet cutoff tends to infinity. Both problems are related to the local point particle nature of the building blocks of the theory. Nonlocality has a chance of removing these problems. Superstring theory, in particular, is nonlocal from the point of view of point particles since the building blocks of the theory are extended objects. The promise of superstring theory for eliminating the divergences of point particle quantum field theories has been long realized (see e.g. \cite{GSW} for an early textbook on string theory). In the following we will sketch a couple of approaches towards developing a string-based early universe cosmology.

\section{Approaches to String Cosmology}

When compared to local point particle effective field theories, strings contain new degrees of freedom and new symmetries. If we imagine space to be a torus of radius $R$ (same radius for all spatial dimensions), then strings have, in addition to the momentum modes (the only modes in point particle theories) whose energies are quantized in units of $1/R$, winding modes whose energies scale with $R$, and an infinite tower of oscillatory modes whose energies are independent of $R$. Associated with these new modes there is a new symmetry, namely the T-duality symmetry which, in this simple setup, is the symmetry under the transformation $R \rightarrow 1/R$ (in string units). Under this symmetry, the role of momentum and winding modes is exchanged. As realized in \cite{BV}, simple thermodynamic arguments indicate that string theory can resolve the temperature singularity of SBB cosmology: consider gradually shrinking a box filled with a gas of strings. Initially the energy is in the momentum modes, the degrees of freedom which are light when $R$ is large. As the box shrinks, the temperature $T$ increases until the point when the string energy densitiy is reached, at which point the string oscillatory modes can be excited. As $R$ shrinks further, then instead of the temperature increasing, more and more different oscillatory modes are excited. Once $R$ shrinks below the string scale, the energy flows into the winding modes, and $T$ decreases. The T-duality symmetry implies
\be
T(R) \, = \, T(1/R) \, .
\ee
As studied in \cite{BV}, the range of values of $R$ for which $T(R)$ hovers near the maximal temperature $T_H$ increases as the entropy of the system increases.

In \cite{BV}, a new model of early universe was proposed based on the above considerations - {\it String Gas Cosmology} (SGC). It is based on the hypothesis that the universe emerges as a hot gas of strings in thermal equilibrium on a compact background space (most easily visualized as a nine-dimensional torus). The hot string gas contains momentum, winding and oscillatory modes. Winding modes prevent space from expanding while momentum modes prevent space from contracting. Without string interactions, all spatial dimensions would remain compactified at the string scale. Winding modes, however, can interact and decay into string loops, but this process cannot liberate more than three spatial dimensions since in more than three large spatial dimensions the string world sheets cannot intersect. Hence, a phase transition in which winding modes decay can render precisely three dimensions of space large \cite{BV}, while the other dimensions are kept compact at the string scale by the joint effects of string winding and momentum modes \cite{moduli}. 

Since the early phase (Hagedorn phase) of SGC is a hot gas of strings, the origin of fluctuations will be thermal and not vacuum. Thermal fluctuations of the hot string gas lead to curvature fluctuations and gravitational waves, the former determined by the energy density perturbations, the latter by the anisotropic pressure inhomogeneities. The matter correlation functions can be computed beginning with the partition function of the string gas, and it was shown in \cite{NBV} that a nearly scale-invariant spectrum of both curvature fluctuations and gravitational waves results. The scale-invariance is due to the holographic scaling of the correlation functions, which in turn is due to the fact that the basic objects are strings and not point particles. The model predicts a slight blue tilt $n_t$ of the spectrum of gravitational waves, while standard canonical inflation models always predict a red tilt. There is a characteristic consistency relation between the predicted red tilt $n_s - 1$ of the curvature fluctuations and the gravitational wave tilt \cite{tilt}
\be
n_t \, = \, 1 - n_s \, .
\ee

The challenge for String Gas Cosmology is to obtain a self-consistent description of the Hagedorn phase. Such a description will clearly have to go beyond an effective field theory analysis. As already argued in \cite{BV}, momentum and winding modes need to be considered on an equal footing in this phase. In point particle theories, the position operator $|x>$ is defined as the Fourier transform of the momentum eigenstates $|n>$
\be
|x> \, = \, \sum_n e^{i n x} |n> \, .
\ee
In string theory, there is a dual position operator ${\tilde{x}}$ defined in terms of the winding eigenstates $|m>$
\be
|{\tilde{x}}> \, = \, \sum_m e^{i m {\tilde{x}}} |m> \, .
\ee
When $R$ is large, then the momentum modes are light and hence an experimenter with a limited amount of energy will measure space in terms of $|x>$, while for small $R$ it is the winding modes which are light and distance will be measured in terms of the dual variables $|{\tilde{x}}>$. At the self-dual radius ($R = 1$ in string units), the original and dual spatial variables must be on equal footing, and this makes it clear that an effective field theory description based on point particles (which only can access the $|x>$ variables) will break down.

 {\it Double Field Theory} (DFT) \cite{Siegel, HZ} is an attempt to construct an effective field theory for the low energy degrees of freedom of string theory which is consistent with the T-duality symmetry. It is a field theory which lives in the double space containing both $d$ $|x>$ and $d$ $|{\tilde{x}}>$ coordinates (where $d$ is the number of the ``usual'' spatial dimensions). The gravitational action involves a {\it generalized metric} on the doubled space which is constructed from the usual metric $g_{\mu \nu}$ of $d + 1$ dimensional space-time and the antisymmetric tensor field $B_{\mu \nu}$ which is part of the low energy sector of string theory. The action is chosen such that, in the case when fields depend only on one set of spatial coordinates, the usual supergravity action for $g_{\mu \nu}, B_{\mu \nu}$ and the dilaton field is recovered. This is the ``vacuum sector'' of the theory.
 
It is possible to couple a perfet fluid matter energy-momentum tensor to the vacuum sector of DFT. This tensor can be chosen to represent a string gas living on the DFT background \cite{us}. With an ansatz of homogeneity and isotropy, the resulting equations of motion can be analyzed, and it can be shown that the solutions are nonsingular when interpreted in terms of physical variables. On the other hand, no quasi-static Hagedorn phase is found. In the case of a cosmological background, Hohm and Zwiebach \cite{HZ2} were able to classify all possible $\alpha^{\prime}$ corrections to the gravitational action, and the analysis was extended in \cite{us2} to allow for the presence of matter, and in \cite{us3} to the case of anisotropic space-times with matter sources.

In \cite{us3} we studied a Bianchi cosmology with 3 external dimensions given by a scale factor $a(t)$, and 6 internal dimensions with scale factor $b(t)$. An ansatz for the equation of state parameters $w_i$ and $w_e$ (where $w$ is the ratio of pressure to energy density, and the subscripts $i$ and $e$ stand for the internal and external dimensions, respectively) was made to represent an initial isotropic string gas which undergoes a phase transition at a time $t$ which we can take to be $t = 0$ when the winding modes about the external directions decay into string loops. Thus, for $t \ll 0$ we have an equation of state of winding modes for all dimensions, i.e. equation of state parameters $w_e = w_i = - 1/9$. After the phase transition the internal dimensions are fixed at the string scale by the interplay between momentum and winding modes, while the equation of state in the three external dimensions is that of radiation. Thus, for $t \gg 0$ we choose $w_i = 0$ and $w_e = 1/3$. Given certain conditions on the corrections terms in the action, solutions were found in which for $t \ll 0$ both scale factors are constant in the Einstein frame, which is what is assumed in SGC. However, the dilaton is time-dependent. For $t \gg 0$ the internal dimensions are static and the external dimensions evolve as in a radiation-dominated Friedmann cosmology (the dilaton is fixed). An open question is how to generalize the approach of \cite{HZ2} to inhomogeneous space-times. If this could be done, it would then be possible to study the generation and evolution of cosmological perturbations.

Current approaches to superstring cosmology, even though they go beyond point particle effective field theories, remain incomplete since they are not based on a complete non-perturbative formulation of string theory. Matrix models \cite{BFSS, IKKT} (see also \cite{Antal}) have been suggested as such a formulation. For example, the BFSS proposal \cite{BFSS}, starts with a quantum mechanical model of nine $N \times N$ Hermitean matrices $X_i, \, i = 1, \cdots , 9$ given by the Lagrangian (focusing only on the bosonic part of the model)
\be
L \, = \, \frac{1}{2 g^2} \bigl[ {\rm Tr} \bigl( \frac{1}{2} (D_t X_i)^2 - \frac{1}{4} [X_i, X_j]^2 \bigr) \bigr] 
\ee
where $D_t$ is a covariant derivative involving a tenth $N \times N$ matrix $X_0$, and $g$ is a coupling constant. In the limit $N \rightarrow \infty$ with $g^2 N$ held fixed the full model (including fermions) yields a non-perturbative definition of M-theory.

In a high temperature state, the bosonic part of the BFSS matrix model is equivalent to the bosonic part of the IKKT matrix model \cite{IKKT}. It is always possible to choose a basis in which the temporal matrix $X_0$ is diagonal. Detailed studies of the full IKKT model (see e.g. \cite{Nishimura} for a recent review) have shown that, in this basis, the spatial matrices become block-diagonal, and that the trace of these matrix blocks can be identified with the extent of space in the corresponding direction. Thus, the matrix model yields emergent time and emergent space. Furthermore, the studies have shown that in the large time limit, the state yields spontaneous $SO(9)$ symmetry breaking, only three of the spatial parameters becoming large. This is reminiscent of what happens in String Gas Cosmology. 

In a recent work \cite{us4}, we have studied thermal fluctuations of the BFSS model in the high temperature state, and computed the resulting cosmological fluctuations and gravitational waves at late times, assuming that a phase transition between the early emergent phase and the phase of Standard Big Bang expansion occurs. We find scale-invariant spectra for both curvature perturbations and gravitational waves, like in String Gas Cosmology \footnote{For another non-perturbative proposal for the emergent state of string cosmology see \cite{Vafa}.}

\section{Discussion}

The current paradigm of early universe cosmology is based on point particle local effective field theory. I have pointed out problems of this approach for early universe cosmology, in particular in the case of an accelerated expansion of space. The focus was on the {\it Trans-Planckian Censorship Conjecture}. Other conceptual problems arise in the context of string theory: only a small subset of possible effective field theories can be consistent with string theory according to the {\it swampland} criteria (see e.g. \cite{swamp} for reviews). This reinforces the conclusion that one must go beyond local effective field theories to provide a consistent theory of the very early universe. Note that there are interesting connections between the swampland criteria and the TCC (see \cite{Bedroya3, Suddho}).

I have described a couple of approaches to early universe cosmology in the context of superstring theory which use nonlocal concepts in a key way, and which give rise to cosmologies which do not contain an inflationary phase. However, it is also possible that one might be able to construct inflationary models based on non-perturbative methods. Recently, concrete models have been suggested in \cite{Dvali} and \cite{Keshav}. In both constructions, there is an upper bound on the number of e-foldings of inflation, and in the case of the model of \cite{Keshav} they are consistent with the TCC bound.

\acknowledgments
The author is supported in part by a NSERC Discovery Grant, and by funds from the Canada Research Chair program. He thanks his collaborators, in particular A. Bedroya, M. Loverde, C. Vafa (work on the TCC), H. Bernardo, R. Costa, G. Franzmann, A. Weltman (work on DFT cosmology), S. Brahma and S. Laliberte (work on matrix theory cosmology). A special thanks to L. Buoninfante and S. Kumar for so wonderfully running this online workshop.

\end{document}